\documentclass[10pt,twocolumn,showpacs,amsmath,amssymb,floatfix,superscriptaddress]{revtex4-2}

\usepackage{graphicx}
\usepackage{dcolumn}
\usepackage{bm}
\usepackage{xcolor}
\usepackage{txfonts}
\usepackage{microtype}
\usepackage{epsfig}
\usepackage{hyperref}
\begin{document}
\title{Signatures of linear Breit-Wheeler pair production in polarized $\gamma\gamma$ collisions}
\author{Qian Zhao}
\affiliation{Ministry of Education Key Laboratory for Nonequilibrium Synthesis and Modulation of Condensed Matter, Shaanxi Province Key Laboratory of Quantum Information and Quantum Optoelectronic Devices, School of Physics, Xi'an Jiaotong University, Xi'an 710049, China}
\author{Liang Tang}
\affiliation{College of Physics and Hebei Key Laboratory of Photophysics Research and Application, Hebei Normal University, Shijiazhuang 050024, China}
\author{Feng Wan}
\affiliation{Ministry of Education Key Laboratory for Nonequilibrium Synthesis and Modulation of Condensed Matter, Shaanxi Province Key Laboratory of Quantum Information and Quantum Optoelectronic Devices, School of Physics, Xi'an Jiaotong University, Xi'an 710049, China}
\author{Bo-Chao Liu}
\affiliation{Ministry of Education Key Laboratory for Nonequilibrium Synthesis and Modulation of Condensed Matter, Shaanxi Province Key Laboratory of Quantum Information and Quantum Optoelectronic Devices, School of Physics, Xi'an Jiaotong University, Xi'an 710049, China}
\author{Ruo-Yu Liu}
\affiliation{School of Astronomy and Space Science, Nanjing University, 210023 Nanjing, Jiangsu, China}
\author{Rui-Zhi Yang}
\affiliation{CAS Key Labrotory for Research in Galaxies and Cosmology, Department of Astronomy, School of Physical Sciences, University of Science and Technology of China, Hefei, Anhui 230026, China}
\author{Jin-Qing Yu}\email{jinqing.yu@hnu.edu.cn}
\affiliation{School of Physics and Electronics, Hunan University, Changsha 410082, China}
\author{Xue-Guang Ren}
\affiliation{Ministry of Education Key Laboratory for Nonequilibrium Synthesis and Modulation of Condensed Matter, Shaanxi Province Key Laboratory of Quantum Information and Quantum Optoelectronic Devices, School of Physics, Xi'an Jiaotong University, Xi'an 710049, China}
\author{Zhong-Feng Xu}
\affiliation{Ministry of Education Key Laboratory for Nonequilibrium Synthesis and Modulation of Condensed Matter, Shaanxi Province Key Laboratory of Quantum Information and Quantum Optoelectronic Devices, School of Physics, Xi'an Jiaotong University, Xi'an 710049, China}
\author{Yong-Tao Zhao}
\affiliation{Ministry of Education Key Laboratory for Nonequilibrium Synthesis and Modulation of Condensed Matter, Shaanxi Province Key Laboratory of Quantum Information and Quantum Optoelectronic Devices, School of Physics, Xi'an Jiaotong University, Xi'an 710049, China}
\author{Yong-Sheng Huang}\email{huangys59@mail.sysu.edu.cn}
\affiliation{School of Science, Shenzhen Campus of Sun Yat-sen University, Shenzhen 518107, P.R. China}
\affiliation{Institute of High Energy Physics, Chinese Academy of Sciences, Beijing 100049, China}
\author{Jian-Xing Li}\email{jianxing@xjtu.edu.cn}
\affiliation{Ministry of Education Key Laboratory for Nonequilibrium Synthesis and Modulation of Condensed Matter, Shaanxi Province Key Laboratory of Quantum Information and Quantum Optoelectronic Devices, School of Physics, Xi'an Jiaotong University, Xi'an 710049, China}

\date{\today}
\begin{abstract}
Laser-driven brilliant controllable polarized $\gamma$-photon sources open the way for designing compact $\gamma\gamma$ collider, which enable the large yield of linear Breit-Wheeler (LBW) pairs in a single shot and thus provide an opportunity for the investigation of polarized LBW process. In this work we investigate the polarization characteristics of LBW pair production via our developed spin-resolved binary collision simulation method. Polarization of $\gamma$-photons  modifies the kinematics of scattering particles   and induces the correlated energy-angle shift of LBW pairs, and the latter's  polarization characteristic depends on the helicity configures of scattering particles.  Our method confirms that the polarized $\gamma\gamma$ collider with an asymmetric setup can be performed with currently achievable laser facilities to produce abundant polarized LBW pairs, fulfilling the detection power of polarimetries. The precise knowledge of polarized LBW process is in favor of the calibration and monitor of polarized $\gamma\gamma$ collider, and could enhance the opacity of $\gamma$-photons in high-energy astrophysical objects to exacerbate the inconsistency between some observations and standard models.
\end{abstract}

\maketitle
Quantum electrodynamics (QED) predicts the interaction between two real photons, which leads to the linear Breit-Wheeler (LBW) electron-positron ($e^\pm$) pair production \cite{Breit1934} and photon-photon elastic scattering \cite{Halpern1933}. Although both the processes are observed in the collision of two virtual photons by means of the equivalent photon approximation of ultra-relativistic heavy-ion beams \cite{nature2017,Adam2021}, the validation via real photon-photon collisions has never been realized due to the lack of high-brilliance $\gamma$-photon beam. One of the most important physical projects in the planned $\gamma\gamma$ collider is to search for higher-order photon-photon scattering, which is the solid evidence of the vacuum polarization and is of interest as a search for new physics \cite{Yamaji2016,Drebot2017study,Takahashi2018}. Because the cross section of the lowest-order LBW process is  several orders of magnitude larger than that of the photon-photon scattering \cite{Drebot2017}, it produces useful signatures for diagnostics, feedback, and luminosity optimization in a $\gamma\gamma$ collider \cite{Pak2004,Esberg2014}. Therefore, it is necessary to investigate the comprehensive physics of the LBW process, especially with polarized photons and energy distribution. Moreover, despite the significance in validating basic QED theory, the LBW process is one of the most elemental ingredients of pair plasma production in high-energy astrophysical environment \cite{Ruffini2010}, such as $\gamma$-ray bursts \cite{Kumar2015}, black hole accretion \cite{Hirotani2016,Akiyama2019} and active galactic nuclei \cite{Bottcher2019}.

Thanks to the developments of $\gamma$-photon sources driven by the electron beams of laser wakefield acceleration \cite{Esarey2009}, high-brilliance $\gamma$-photon beams with MeV energy can be produced experimentally through bremsstrahlung \cite{Glinec2005,Lemos2018},  nonlinear Thomson scattering \cite{Sarri2014,Yan2017} and inverse Compton scattering \cite{Yu2016,Cole2018, Zhu2020}, which open the way to investigate the real photon-photon interaction. Recently, numerous theoretical proposals on the $\gamma\gamma$ collider are put forward to validate the LBW process \cite{Pike2014,Ribeyre2016,Drebot2017,Jansen2018,Yu2019,Wang2020comm,Wang2020,Golub2021,Esnault2021,Kettle2021}.
Among those proposals,  the $\gamma\gamma$ collider is designed either with GeV-energy photons from bremsstrahlung inside a high-Z target and keV-energy partner photon from laser-target radiation or X-ray free-electron laser \cite{Pike2014,Golub2021,Kettle2021}, or with two same MeV $\gamma$-photon sources from laser-driven synchrotron or nonlinear Compton scattering \cite{Ribeyre2016,Drebot2017,Jansen2018,Yu2019,Wang2020}. Moreover, the dominated LBW process is studied in laser-driven plasmas \cite{He2021,HeYT2021}, and the quasiparticle-hole pair production in gapped graphene monolayers can be analogous to the LBW process \cite{Golub2020}. However, above proposals mainly focus attention on the large pair yield, but generally ignore the inherent spin effects of scattering particles. The polarization transfer (i.e., helicity transfer) between initial photons and final $e^\pm$ pairs is associated with their spin angular momentum.
Based on the production of high-brilliance photons with highly circular polarization \cite{Yakimenko2006,Omori2006,Alexander2008, Abbott2016,Yan2019,LiYF2020}, the LBW process can be investigated in a polarized $\gamma\gamma$ collider.
  The fundamental physics of helicity in the LBW process has been analyzed qualitatively from the perspective of angular momentum conservation \cite{Baier2002}.
The latest experiment confirms that the linearly polarized photons can induce the azimuthal-angle distribution of the LBW pairs \cite{Adam2021}, since the linear polarization is related to the azimuthal angle of pair momenta. In addition, the impact of energy distribution of $\gamma$-photon beam on the LBW pair yield is analyzed in a semi-analytical model \cite{Esnault2021}. However, considering realistic polarization and energy distribution into laser-driven $\gamma$-photon beams, the spin-associated momentum and polarization characteristics of the LBW process have not been uncovered and are still great challenges.

\begin{figure}[!t]
\setlength{\abovecaptionskip}{0.2cm}  	
\centering\includegraphics[width=1\linewidth]{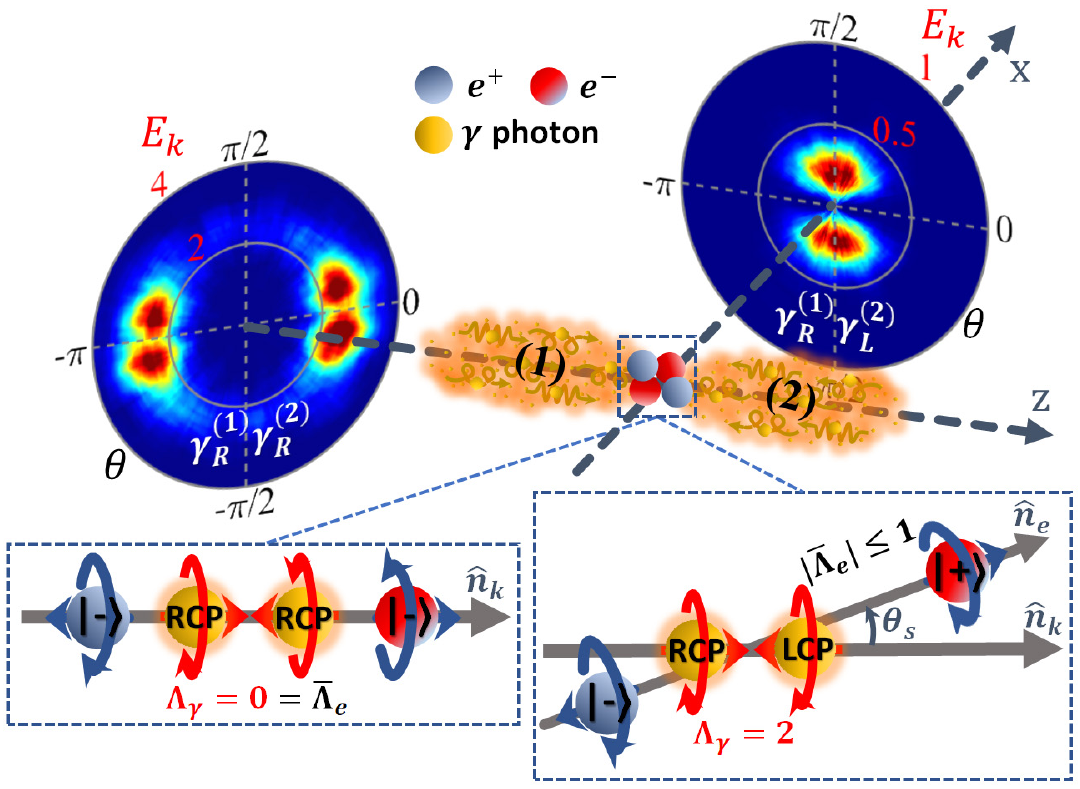}
\caption{Scenario of the LBW pair production in a colliding setup of polarized $\gamma$-photon beams in $x-z$ plane in the laboratory frame. The pair spectra are presented in the plane of polar angle $\theta$ and kinematic energy $E_k$. The lower sketches show the helicity transfer of the LBW process in the center of mass (c.m.) frame, and $\Lambda_\gamma$ is the total helicity of two-photon system in the direction of photon momentum $\hat{n}_k$. The scattered pair carries the mean total helicity $\overline{\Lambda}_e$ in the direction of electron momentum $\hat{n}_e$,  $|\pm\rangle$ represent the positive and negative helicity states, corresponding to the right-hand and left-hand spirals, respectively, and  $\gamma_{R/L}^{(1)/(2)}$ denote right-hand circular polarization (RCP) or left-hand circular polarization (LCP) photon in beam (1) and beam (2), respectively.}
	\label{fig1}
\end{figure}

\begin{figure}[!t]
\setlength{\abovecaptionskip}{0.2cm}  	
\centering\includegraphics[width=1\linewidth]{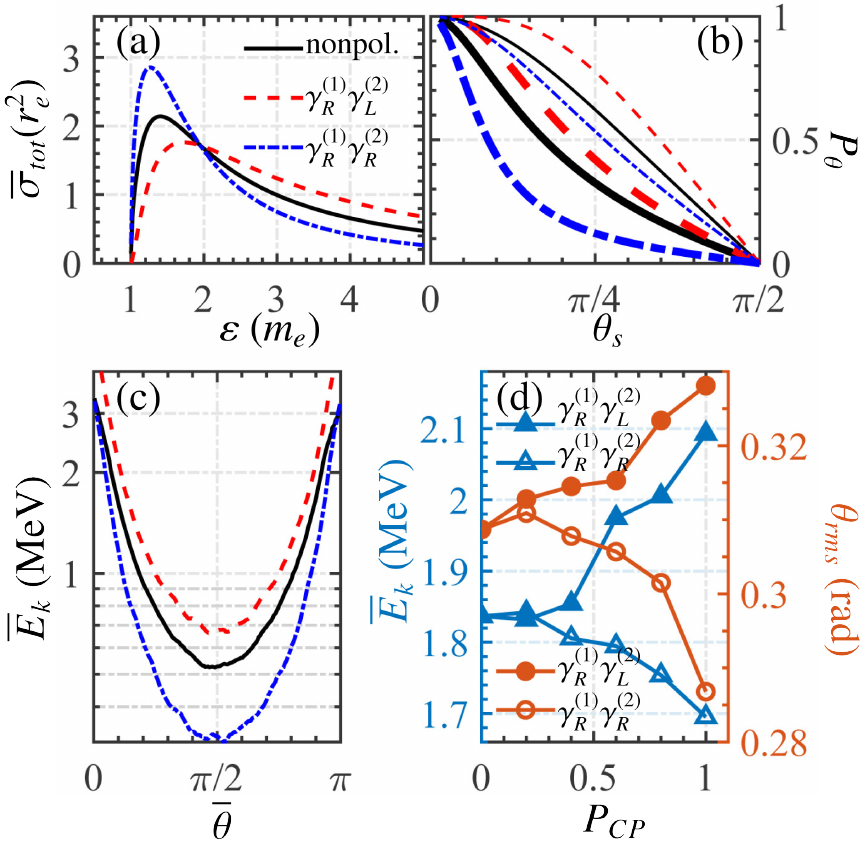}
\caption{(a) $\overline{\sigma}_{tot}$ vs the c.m. energy $\varepsilon$ for different  collision schemes. The black-solid, red-dashed and blue-dash-dotted lines in (a) and (b) indicate the cases of employing nonpolarized, $\gamma_{R}^{(1)}\gamma_{L}^{(2)}$ and $\gamma_{R}^{(1)}\gamma_{R}^{(2)}$ $\gamma$-photons, respectively.  (b) $P_\theta=\overline{\sigma}_\theta/\overline{\sigma}_{tot}$ vs $\theta_s$. The thin and thick lines correspond to  $\varepsilon$ = 1.4 and 4, respectively.
	(c) Average kinetic energy $\overline{E}_k$ vs average polar angle $\overline{\theta}$ of positrons. (d) $\overline{E}_k$ and divergence angle $\theta_{rms}$ (root-mean-square deviation) of positrons beamed into $0<\theta<\pi/6$ vs average circular polarization $P_{CP}$ of the initial $\gamma$-photon beam. The results in (c) and (d) are simulated with colliding $\gamma$-photon beams with an exponential energy distribution at an average energy 2 MeV and a divergence angle 0.1 rad in the laboratory frame.}
\label{fig2}
\end{figure}

\begin{figure}[!t]
\setlength{\abovecaptionskip}{0.2cm}  	
\centering\includegraphics[width=1\linewidth]{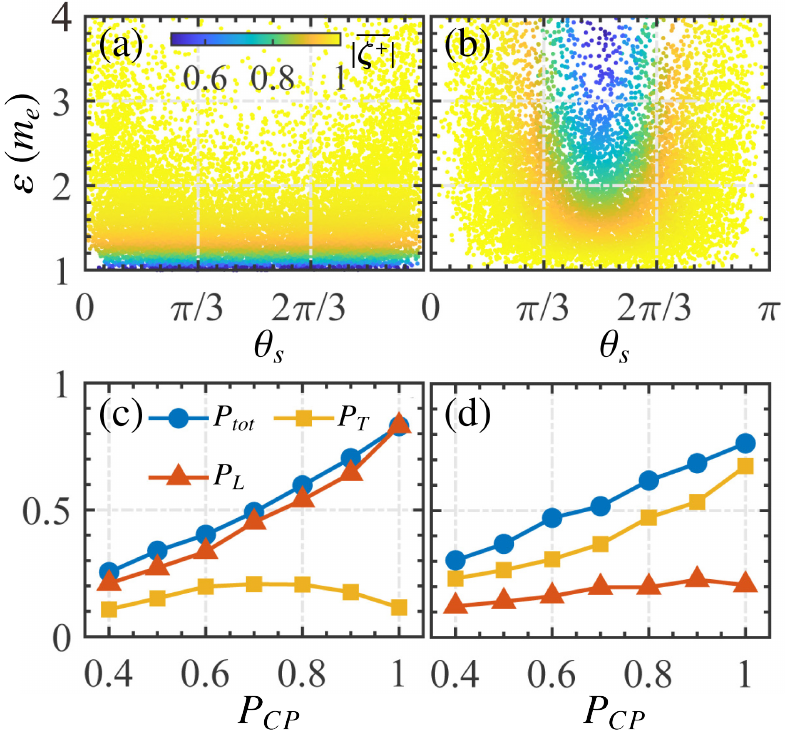}
\caption{
(a) and (b): Distributions of  $|\overline{\bm{\zeta}^{+}}|$ of produced positrons for the $\gamma_R^{(1)}\gamma_R^{(2)}$ and $\gamma_R^{(1)}\gamma_L^{(2)}$ collisions, respectively. (c) and (d):  Variations of  $P_{\rm{tot}}$, $P_L$ and  $P_T$ of positrons with respect to $P_{CP}$ of the $\gamma$-photon beam,  extracted from positrons within  $0<\theta<\pi/6$ for the $\gamma_R^{(1)}\gamma_R^{(2)}$ collision and $\pi/3<\theta<2\pi/3$ for the $\gamma_R^{(1)}\gamma_L^{(2)}$ collision, respectively. Here employed $\gamma$-photon beams both have an uniform energy distribution between 0.1 MeV and 2 MeV. }
	\label{fig3}
\end{figure}

In this Letter, we investigate the complete polarization effects in the LBW process by virtue of our first developed spin-resolved Monte Carlo (MC) simulation method, which is applicable for general binary collisions of leptons and photons (see the interaction scenario in Fig.~\ref{fig1}).
We find that the circular polarization of $\gamma$-photons  can modify the kinematics of scattering particles   and induces a correlated energy-angle shift of the LBW pairs (see the upper panel in Fig.~\ref{fig1} and more details in Fig.~\ref{fig2}), and the polarization characteristic of the LBW pairs
depends on the helicity configures of scattering particles  (see lower panel in Fig.~\ref{fig1} and more details in Fig.~\ref{fig3}).
Our method confirms that the polarized $\gamma\gamma$ collider with an asymmetric setup can be performed with currently achievable laser facilities (see Tab.~\ref{Tab:beam}),
and  the considered polarization effects may have significant applications in  high-energy astrophysics.

Let us first summarize our simulation method for calculating the production and polarization of the LBW pairs.  Employing the standard treatment of particle density matrixes in the scattering amplitude of incoherent binary collisions  \cite{Berestetskii1982,Kotkin1998,Grozin2002,Baier2002,Ivanov2004,Ivanov2005}, the polarized LBW cross section is obtained as
\begin{align}
  \frac{{\rm d}\sigma}{{\rm d}\Omega}= \sigma_0\left[F+\sum_{i=1}^{3}(G_i^+\zeta_i^++G_i^-\zeta_i^-)+\sum_{i,j=1}^{3}H_{i,j}\zeta_i^+\zeta_j^-\right],\label{pairproduction}
\end{align}
where $\sigma_0=r_e^2m_e^2|\bm{p}_e|/16\varepsilon^3$, $\Omega$ is the solid angle,  $\bm{p}_e$ the c.m. momentum of electron, $\varepsilon$  the c.m. energy of photon, electron and positron (they are equal in the c.m. frame), $m_e$ and $r_e$ the electron mass and classical radius, respectively,  $\zeta^\pm_i$ the spin components of electron (``$-$'') and positron (``$+$''), and   the factors $F$, $G_i^{\pm}$ and $H_{i,j}$ include the photon Stokes parameters $\xi_i$ \cite{Berestetskii1982} and are given  in  \cite{supplement}.
Summing over $\zeta^\pm_i$   and integrating over $\Omega$, one  obtains
\begin{align}\label{sigt}
\overline{\sigma}_{tot}&=\frac{r_e^2m_e^4\pi}{4\varepsilon^4}\sqrt{\frac{s-4}{s}}\left( -s -4+2\xi _1^{(1)}\xi _1^{(2)}+3s \xi _2^{(1)}\xi _2^{(2)}\right.\nonumber\\
&-\left.2 \xi _3^{(1)}\xi _3^{(2)}\right)+\frac{16\pi}{s}\tanh^{-1}{\sqrt{\frac{s-4}{s}}}\left(2s^2+8s-16\right.\nonumber\\
&+\left.8 \xi _1^{(1)}\xi _1^{(2)}-2 s^2\xi _2^{(1)}\xi _2^{(2)}-8\xi _3^{(1)}\xi _3^{(2)}\right),
\end{align}
where $s=4\varepsilon^2/m_e^2$.  Relativistic units with $c=\hbar=1$ are used throughout.

The event probabilities are given by $\overline{\sigma}_{tot}$ and the pair production is determined by a rejection method.
The colliding photons in each 3-dimensional cell are paired in the laboratory frame using the no-time-counter method \cite{Gaudio2020}. By Lorentz boost along the c.m. frame velocity $\bm{\beta}_{cm}$
 \cite{Schlickeiser1997,Ribeyre2017,Ribeyre2018}, the 4-momenta of paired photons are transformed into the c.m. frame to calculate $\bm{p}_e$, which  is determined by the probabilistic scattering angle $\theta_s$. Here $\theta_s$ is calculated by solving $\overline{\sigma}_\theta/\overline{\sigma}_{tot}=R_1 \in(-1,1)$, where $R_1$ is an uniform random number and $\overline{\sigma}_\theta=\sqrt{s(s-4)/4}\int_{-|\cos{\theta_s}|}^{|\cos{\theta_s}|}{\rm d}\overline{\sigma}$. Moreover, the energy and momenta of $e^\pm$ in the laboratory frame are obtained by the inverse Lorentz boost acting on $\varepsilon$ and $\bm{p}_e$.

 The mean spin polarization vectors of electron and positron $\overline{\bm{\zeta}^{\pm}}$ are determined by the defined 3-vector basis \cite{Kotkin1998, supplement} and their components can be analytically  calculated
 via  $\overline{\zeta^{\pm}}_{i}=G_i^\pm/F$ \cite{Berestetskii1982}. The corresponding mean helicities are $\lambda_\pm=\mp\overline{\bm{\zeta}^{\pm}}\bm{p}_e/2|\bm{p}_e|$. In our MC method,
the projections of  $\overline{\bm{\zeta}^{\pm}}$ onto the defined spin states $\bm{D}^{\pm}$ (unit vectors) of a detector are calculated with transition probabilities, and the latter determine the sign of components $D^\pm_{i}=\overline{\zeta^{\pm}}_{i}/|\overline{\bm{\zeta}^{\pm}}|$. Consequently, the total beam polarization is calculated by $P_{tot}=\sqrt{(\overline{D^{\pm}_1})^2+(\overline{D_2^{\pm}})^2+(\overline{D_3^{\pm}})^2}$, where $\overline{D^\pm_{i}}=\sum_{n=1}^{N}
D^{\pm,n}_{i}/N$  with the particle number $N$ \cite{Tolhoek1956}. Aligning $\bm{D}^\pm$ to the parallel or perpendicular direction of the momenta can further obtain the longitudinal polarization $P_L$ or transverse  polarization $P_T$ for $e^\pm$ beams. More details of our  simulation method are clarified in the Supplementary Materials \cite{supplement}.

 Impact of the $\gamma$-photon polarization on the energy and polar angle distributions of the LBW pairs is shown in Fig.~\ref{fig2}.
$\overline{\sigma}_{\rm{tot}}$ of the $\gamma_{R}^{(1)}\gamma_{R}^{(2)}$ case  increases about $33\%$ at the peak energy and narrows the spectrum, compared with that of the nonpolarized case [see Fig.~\ref{fig2}(a)], while the linear polarization only modifies the amplitude because it has no impact on $\theta_s$ \cite{supplement}. A definite-shaped $P_\theta(\theta_s)$ implies that the produced electron (positron) is scattered into a certain range of d$\theta_s$ with a corresponding probability of d$P_\theta$ [see Fig.~\ref{fig2}(b)]. The reactions within $\theta_s\lesssim\pi/6$ in the $\gamma_R^{(1)}\gamma_L^{(2)}$ collision near threshold energy are almost forbidden due to the approximate zero probabilities, which lead to a dipole angular spectrum around the perpendicular direction of the colliding axis. By contrast,  the dominated reactions occur within $\theta_s\lesssim\pi/6$ in the $\gamma_R^{(1)}\gamma_R^{(2)}$ collision at the energy much beyond the threshold, which lead to a quadrupole  angular spectrum around the colliding axis (see energy-angle spectra Fig.~\ref{fig1}). The distinct energy-correlated $\theta_s$ in these two interaction schemes causes the energy-angle correlation of the pairs, whereby the $\gamma_{R}^{(1)}\gamma_{L}^{(2)}$ collision results in the larger kinetic energy $\overline{E}_k$ than that of the $\gamma_{R}^{(1)}\gamma_{R}^{(2)}$ collision [see Fig.~\ref{fig2}(c)]. As the initial circular polarization ($P_{CP}$) of $\gamma$-photons increases, in the $\gamma_{R}^{(1)}\gamma_{L}^{(2)}$ collision  the average kinetic energy $\overline{E}_k$ and the divergence angle $\theta_{rms}$ of positrons both increase as well, while in the $\gamma_{R}^{(1)}\gamma_{R}^{(2)}$ collision the tendency is inverse   [see Fig.~\ref{fig2}(d)]. Thus, $\theta_{rms}$ is positively correlated to $\overline{E}_k$. The influence of the energy and divergence-angle fluctuations of the $\gamma$-photon beams on $\theta_{rms}$ and $\overline{E}_k$ is estimated in \cite{supplement} and uniform results can be obtained with flexible parameters. Note that the correlated polar angle and energy distributions of the LBW pairs can be resolved precisely in experiments by the single particle detector \cite{Kettle2021}.

\begin{figure}[!t]
\setlength{\abovecaptionskip}{0.2cm}  	
\centering \includegraphics[width=1\linewidth]{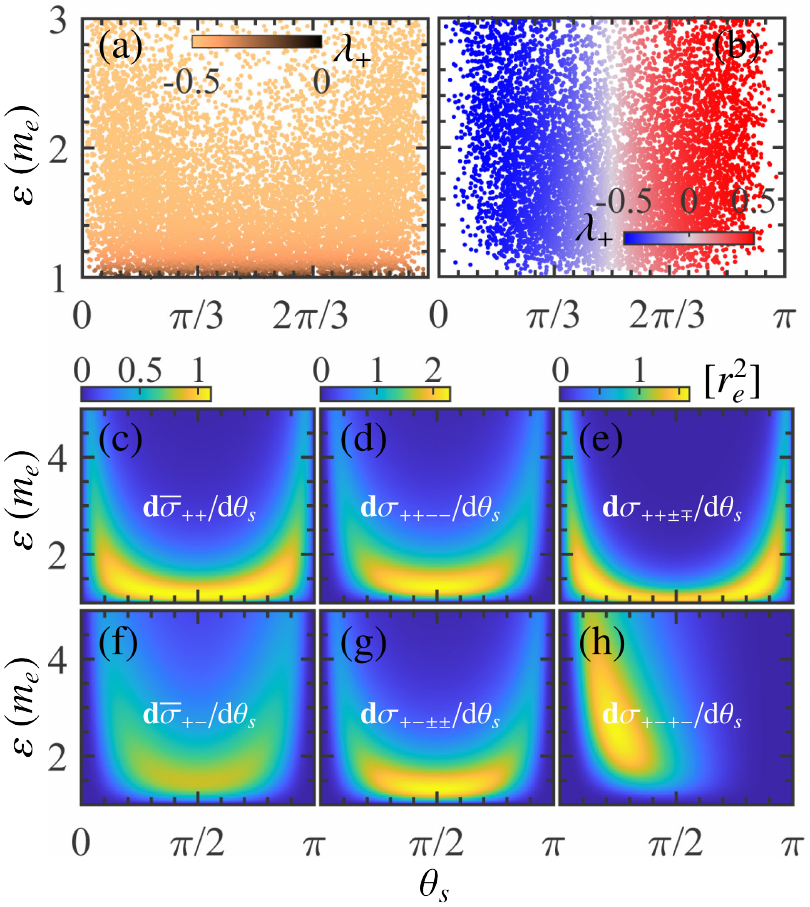}
\caption{(a) and (b): Distributions of the positron helicity $\lambda_+$  with respect to $\theta_s$ and $\varepsilon$ in the c.m. frame for the $\gamma_{R}^{(1)}\gamma_{R}^{(2)}$  and $\gamma_{R}^{(1)}\gamma_{L}^{(2)}$ collisions, respectively. Employed parameters are the same with those in Fig.~\ref{fig3}.
(c)-(e) [(f)-(h)]: Differential cross sections for different helicity channels for the  $\gamma_{R}^{(1)}\gamma_{R}^{(2)}$ ($\gamma_{R}^{(1)}\gamma_{L}^{(2)}$) collision. Here   the subscripts from the first to the fourth in sequence denote positive (``+'') or negative (``-'') helicity eigenstates of $\gamma^{(1)}$, $\gamma^{(2)}$, $e^-$, and $e^+$, and d$\overline{\sigma}$  indicates the spin-summarized cross section and is calculated via $\sigma_0 F$ in Eq.~(\ref{pairproduction}).}
	\label{fig4}
\end{figure}

Particularly, the spin-polarization of the LBW pairs is derived from  the circular polarization of parent $\gamma$-photons.
For the $\gamma_R^{(1)}\gamma_R^{(2)}$ collision,  the partial polarization is produced near the threshold energy of the pair production [see Fig.~\ref{fig3}(a)], while for the $\gamma_R^{(1)}\gamma_L^{(2)}$ collision,  the partial polarization is produced around $\theta_s=\pi/2$  [see Fig.~\ref{fig3}(b)]. It is nontrivial to reveal the variation of the positron polarization with respect to the $\gamma$-photon polarization. For the $\gamma_R^{(1)}\gamma_R^{(2)}$ collision, $P_L$ dominates the polarization and increases linearly, and $P_T$  [attributed to the d$\sigma_{+-\pm\mp}$ channel; see Fig.~\ref{fig4}(e)] is less than 0.2 as $P_{CP}$ varies [see Fig.~\ref{fig3}(c)],  since  the produced pairs possess sole negative helicities [see Fig.~\ref{fig4}(a)]. While, for the $\gamma_R^{(1)}\gamma_L^{(2)}$ collision  $P_T$ dominates the polarization [see Fig.~\ref{fig3}(d)], since the produced pairs possess the mixed helicity states \cite{Hikasa1986} around $\theta_s=\pi/2$ and the magnitude of $P_T$ is affected by the extracted polar angle range [see Fig.~\ref{fig4}(b)].
Thus, one could observe the signatures of polarization in the LBW process either via detecting $P_L$ in the $\gamma_R^{(1)}\gamma_R^{(2)}$ collision around the colliding axis or via detecting $P_T$ in the $\gamma_R^{(1)}\gamma_L^{(2)}$ collision around the perpendicular direction of the colliding axis.

The physical mechanism of the helicity transfer in the LBW process is analyzed in Fig.~\ref{fig4}.
The polarization is derived from the mean-helicity distribution, and the $\gamma_{R}^{(1)}\gamma_{R}^{(2)}$ collision leads to the energy-dependent negative helicities, while the $\gamma_{R}^{(1)}\gamma_{L}^{(2)}$ collision leads to the angle-dependent alternating helicities (vary between -0.5 and 0.5) [see Figs.~\ref{fig4}(a) and (b)]. The positron helicity  originates from the superposition of various helicity eigenstates with different weights determined by the differential cross section. There are only three non-vanished helicity channels d$\sigma_{++--}$, d$\sigma_{+++-}$ and d$\sigma_{++-+}$ for the $\gamma_{R}^{(1)}\gamma_{R}^{(2)}$ collision [see Figs.~\ref{fig4}(d) and (e)].  Here, d$\sigma_{++\pm\mp}$ dominate the reaction near the  threshold energy, produce the positrons at the helicity states $|+\rangle$ or $|-\rangle$ with the same weights, and consequently, cancel each other (i.e., $P_L$ vanishes and $P_T$ maximizes). Thus, only the channel of d$\sigma_{++--}$ contributes to the state $|-\rangle$, i.e., d$\sigma_{++--}$ solely induces $P_L$. These helicity channels finally result in the mean helicity distribution in Fig.~\ref{fig4}(a) and the corresponding positron polarization in Fig.~\ref{fig3}(a). Similarly, in the $\gamma_{R}^{(1)}\gamma_{L}^{(2)}$ collision, there are four helicity channels d$\sigma_{+-\pm\pm}$ and d$\sigma_{+-\pm\mp}$ (note that here d$\sigma_{+--+}$ is symmetric about $\theta_s=\pi/2$ with d$\sigma_{+-+-}$ and thus is not shown) [see Figs.~\ref{fig4}(g) and (h)].
d$\sigma_{+-\pm\pm}$ only contribute to $P_T$ and
d$\sigma_{+-\pm\mp}$ mainly to $P_L$ [see Figs.~\ref{fig4}(b) and \ref{fig3}(b)].

\begin{table}[!t]
\caption{Average current $I$ and polarization (Pol.) of produced positrons (electrons) extracted from the polar angle range of $\Delta\theta$. The symmetric setup includes photons number $N_\gamma$,  brilliance $\mathcal{B}_\gamma   (\rm{photons~s^{-1}mm^{-2}mrad^{-2} 0.1\% BW})$, and  cross angle of  colliding $\gamma$-photon beams. And the asymmetric setup considers a RCP $\gamma$-photon beam with 80\% polarization colliding with a X-ray beam with uniform density $n_{X}$. Employed $\gamma$-photon beam in the symmetric (asymmetric) setup has a $\sim80$ fs duration and an exponential energy distribution averaged at 2 (100) MeV.}
\label{Tab:beam}
\begin{center}
\begin{tabular}{c|c|ccc}
  \hline\hline
  beam parameters&collisions&$I$ [mA] &Pol. &$\Delta\theta$[rad] \\[0.2 ex]
  \cline{1-5}
Symmetric Setup:& \\ 
  \raisebox{0.1em}[0pt]{$N_\gamma:1\times10^{11}$ Phots.}&\raisebox{0.2em}[0pt]{$\gamma_R^{(1)}\gamma_R^{(2)}$} &4.231& $P_L$&$0\pm\pi/6$\\[0.2 ex]
    \cline{2-5}
 \raisebox{0.0em}[0pt]{$\mathcal{B}_\gamma:1.5\times10^{23}$}&\\
\raisebox{0.1em}[0pt]{cross-angle: $5^{\rm{o}}$}&\raisebox{0.2em}[0pt]{$\gamma_R^{(1)}\gamma_L^{(2)}$}&3.952 &$P_T$&$\pi/2\pm\pi/6$ \\[0.2 ex]
 \cline{1-5}
Asymmetric Setup:& \\ 
   \raisebox{0.1em}[0pt]{$N_\gamma:1\times10^{7}$ Phots.}&$\gamma_R^{(1)}(0.8)$ &1.052 & $P_L$(0.348),&$0\pm0.02$\\
  \raisebox{0.1em}[0pt]{$\mathcal{B}_\gamma$: $2.2\times10^{21}$}&\\
\raisebox{0.1em}[0pt]{$n_{X}:9.3\times10^{23}$ cm$^{-3}$}&\raisebox{0.8em}[0pt]{X-ray}&&\raisebox{0.8em}[0pt]{$P_T$(0.151)}\\
  \hline\hline
\end{tabular}
\end{center}
\end{table}

For the experimental feasibility, the symmetric and asymmetric setups are considered to design the polarized $\gamma\gamma$ collider, as shown in Table \ref{Tab:beam}. The symmetric setup reveals the polarization-associated signatures in both  of momentum and spin (see Figs. \ref{fig2} and \ref{fig3}), and the required minimal brilliance of the $\gamma$-photon beam is rather high in order to produce a resolvable mA current for the polarimetry.
In the asymmetric setup, the RCP $\gamma$-photon beam injects into a nonpolarized X-ray bath with an uniform energy distribution between $1-3$ keV \cite{Kettle2021}, produce a strongly collimated positron (electron) beam with  moderate $P_L$ and $P_T$. This asymmetric setup significantly reduces the required minimal brilliance of the $\gamma$-photon beam which is experimentally feasible  through inverse Compton scattering \cite{Albert2010,Yu2016}, and the polarized one is also available by bremsstrahlung radiation with $\sim10^{8}$ photons \cite{Pike2014,Abbott2016, Yan2019}. The yield and polarization stabilities are estimated when the divergence angle and the polarization of the $\gamma$-photon beam change and uniform results are obtained (see \cite{supplement}). In terms of the polarization detection, the $e^\pm$ polarimetry technology has achieved a high precision $\lesssim1\%$, e.g., Compton transmission polarimetry \cite{Omori2006,Alexander2008,Abbott2016} and Mott polarimetry \cite{Tioukine2011,Aulenbacher2018,Grames2020}, which are both applicable at the energy of $0.1-10$ MeV and can response the current as low as $\lesssim100\mu$A \cite{Tioukine2011,Aulenbacher2018}.

Furthermore, we underline that the LBW process is widely involved in high-energy astrophysical phenonema. The measured $\gamma$-ray spectra of those intense compact astrophysical objects (such as GRBs, blazars, and etc) are supposed to be attenuated at the high-energy end by the low-energy radiation therein via this process \cite{Granot2008, Guetta2011, Poutanen2010, Xue2019}. Both the low-energy radiation and the $\gamma$-ray radiation, which generally arise from the synchrotron radiation and the inverse Compton scattering respectively, can be polarized in the presence of magnetic field. Taking into account the polarized LBW process would enhance the opacity of $\gamma$-photons of those sources and consequently exacerbate the inconsistency between some observations and standard models, which may challenge the current understanding on the astrophysical objects. One example is the discovery of minute-scale $\gamma$-ray variability in many blazars \cite[e.g.][]{Aharonian2007,Arlen2012,Meyer2019}. The fast variability implies a compact radiation zone so that an unrealistically large Doppler factor with $\delta_D> 50$ has to be invoked in order to avoid significant attenuation of the $\gamma$-ray flux \cite{Begelman2008}, under the current standard picture of the blazar's jet. The value is significantly larger than the Doppler factors inferred from superluminal motions of blazar jets as revealed by radio observations, which is typically around 10 \cite{Hovatta2009}. This is referred to as the ``Doppler factor crisis'' \cite{Bottcher2019} and has important implications for the internal structure of the blazar jet \citep{Giannios2009}, the external gaseous environment \cite{Tavani2015} or the acceleration mechanism of the relativistic outflow driven by supermassive black hole \cite{Lyutikov2010}. Besides, considering the polarized LBW process could be quite helpful to reveal the radiation mechanism of a pulsar, since the density distribution of the pair plasma in the magnetosphere of a pulsar is distorted by the polarized effect  \cite{Huang2015}.

In conclusion, we develop a fully spin-resolved simulation method for general binary collisions to investigate the complete polarization effects of the LBW process in polarized $\gamma\gamma$ collision. Qualitative signatures of polarized LBW process are imprinted on  momentum and spin of produced pairs.
The precise physics of polarized LBW process is effective to calibrate and monitor the upcoming polarized $\gamma\gamma$ collider, and paves the way for proceeding the elusive photon-photon scattering. Besides,  the polarization-induced fluctuations of the $e^\pm$ density in high-energy astrophysical objects possibly associate with certain of significant observations, which calls for the further investigation.\\

 {\it Acknowledgement:} This work is supported by the National Natural Science Foundation of China (Grants Nos. 12022506, 11874295, 11875219, 11905169,  12175058), the China Postdoctoral Science Foundation (Grant No. 2020M683447), the Natural Science Foundation of Hunan Province, China (Grant No. 2020JJ5031),  the project of Science and Technology on plasma physics Laboratory (No. 6142A04190111), the Innovation Project
of IHEP (542017IHEPZZBS11820, 542018IHEPZZBS12427), and the CAS Center for Excellence in Particle Physics
(CCEPP).

\bibliography{refs}

\end{document}